\documentclass[nofootinbib,superscriptaddress,a4paper,twocolumn,longbibliography]{revtex4-1}
\usepackage{geometry}
\usepackage[bottom]{footmisc}
\usepackage{array,mathtools,amssymb,booktabs}

\newcolumntype{C}{>{$}c<{$}}
\AtBeginDocument{
\heavyrulewidth=.08em
\lightrulewidth=.05em
\cmidrulewidth=.03em
\belowrulesep=.65ex
\belowbottomsep=0pt
\aboverulesep=.4ex
\abovetopsep=0pt
\cmidrulesep=\doublerulesep
\cmidrulekern=.5em
\defaultaddspace=.5em
}

\usepackage{hyperref}
\usepackage[many]{tcolorbox}% version 2.80 (2014/03/31)

\usepackage{amsthm}
\newtheoremstyle{break}% name
  {}%         Space above, empty = `usual value'
  {}%         Space below
  {\itshape}% Body font
  {}%         Indent amount (empty = no indent, \parindent = para indent)
  {\bfseries}% Thm head font
  {.}%        Punctuation after thm head
  {\newline}% Space after thm head: \newline = linebreak
  {}%         Thm head spec

\theoremstyle{definition}

\theoremstyle{break}

\NewTColorBox{NewBoxIII}{ s O{!htbp} }{%
floatplacement={#2},
IfBooleanTF={#1}{float*,width=\textwidth}{float},
colback=yellow!5!white,colframe=yellow!75!black,title=Survey Hypothetical Responses in Class III: Agent of Understanding,
}  

\usepackage[english]{babel}
\usepackage[latin1]{inputenc}
\usepackage{hyperref}
\usepackage{amsmath, amsfonts, mathrsfs}
\usepackage{graphicx}
\usepackage{xspace}
\usepackage{bbm}
\usepackage{natbib}
\usepackage{multirow}
\usepackage{csquotes}
\usepackage{tabulary}
\usepackage{enumerate}
\usepackage[version=4]{mhchem}
\newcommand\mydots{\hbox to 1em{.\hss.\hss.}}

\begin{document}

\title{On scientific understanding with artificial intelligence}

\author{Mario Krenn}
\email{mario.krenn@mpl.mpg.de}
\affiliation{Max Planck Institute for the Science of Light (MPL), Erlangen, Germany.}
\affiliation{Chemical Physics Theory Group, Department of Chemistry, University of Toronto, Canada.}
\affiliation{Department of Computer Science, University of Toronto, Canada.}
\affiliation{Vector Institute for Artificial Intelligence, Toronto, Canada.}
\author{Robert Pollice}
\affiliation{Chemical Physics Theory Group, Department of Chemistry, University of Toronto, Canada.}
\affiliation{Department of Computer Science, University of Toronto, Canada.}
\author{Si Yue Guo}
\affiliation{Chemical Physics Theory Group, Department of Chemistry, University of Toronto, Canada.}
\author{Matteo Aldeghi}
\affiliation{Chemical Physics Theory Group, Department of Chemistry, University of Toronto, Canada.}
\affiliation{Department of Computer Science, University of Toronto, Canada.}
\affiliation{Vector Institute for Artificial Intelligence, Toronto, Canada.}
\author{Alba Cervera-Lierta}
\affiliation{Chemical Physics Theory Group, Department of Chemistry, University of Toronto, Canada.}
\affiliation{Department of Computer Science, University of Toronto, Canada.}
\author{Pascal Friederich}
\affiliation{Chemical Physics Theory Group, Department of Chemistry, University of Toronto, Canada.}
\affiliation{Department of Computer Science, University of Toronto, Canada.}
\affiliation{Institute of Nanotechnology, Karlsruhe Institute of Technology, Eggenstein-Leopoldshafen, Germany.}
\author{Gabriel dos Passos Gomes}
\affiliation{Chemical Physics Theory Group, Department of Chemistry, University of Toronto, Canada.}
\affiliation{Department of Computer Science, University of Toronto, Canada.}
\author{Florian H\"ase}
\affiliation{Chemical Physics Theory Group, Department of Chemistry, University of Toronto, Canada.}
\affiliation{Department of Computer Science, University of Toronto, Canada.}
\affiliation{Vector Institute for Artificial Intelligence, Toronto, Canada.}
\affiliation{Department of Chemistry and Chemical Biology, Harvard University, Cambridge, USA.}
\author{Adrian Jinich}
\affiliation{Division of Infectious Diseases, Weill Department of Medicine, Weill-Cornell Medical College, New York, USA.}
\author{AkshatKumar Nigam}
\affiliation{Chemical Physics Theory Group, Department of Chemistry, University of Toronto, Canada.}
\affiliation{Department of Computer Science, University of Toronto, Canada.}
\author{Zhenpeng Yao}
\affiliation{Chemical Physics Theory Group, Department of Chemistry, University of Toronto, Canada.}
\affiliation{Center of Hydrogen Science, Shanghai Jiao Tong University, 800 Dongchuan Road, Shanghai 200240, China.}
\affiliation{The State Key Laboratory of Metal Matrix Composites, School of Materials Science and Engineering, Shanghai Jiao Tong University, 800 Dongchuan Road, Shanghai 200240, China.}
\affiliation{Innovation Center for Future Materials, Zhangjiang Institute for Advanced Study, Shanghai Jiao Tong University, 429 Zhangheng Road, Shanghai 201203, China.}
\author{Al\'an Aspuru-Guzik}
\email{alan@aspuru.com}
\affiliation{Chemical Physics Theory Group, Department of Chemistry, University of Toronto, Canada.}
\affiliation{Department of Computer Science, University of Toronto, Canada.}
\affiliation{Vector Institute for Artificial Intelligence, Toronto, Canada.}
\affiliation{Canadian Institute for Advanced Research (CIFAR) Lebovic Fellow, Toronto, Canada.}

\begin{abstract}
Imagine an oracle that correctly predicts the outcome of every particle physics experiment, the products of every chemical reaction, or the function of every protein. Such an oracle would revolutionize science and technology as we know them. However, as scientists, we would not be satisfied with the oracle itself. We want more. We want to comprehend how the oracle conceived these predictions. This feat, denoted as scientific understanding, has frequently been recognized as the essential aim of science. Now, the ever-growing power of computers and artificial intelligence poses one ultimate question: How can advanced artificial systems contribute to scientific understanding or achieve it autonomously?

We are convinced that this is not a mere technical question but lies at the core of science. Therefore, here we set out to answer where we are and where we can go from here. We first seek advice from the philosophy of science to \textit{understand scientific understanding}. Then we review the current state of the art, both from literature and by collecting dozens of anecdotes from scientists about how they acquired new conceptual understanding with the help of computers. Those combined insights help us to define three dimensions of android-assisted scientific understanding: The android as a I) computational microscope, II) resource of inspiration and the ultimate, not yet existent III) agent of understanding. For each dimension, we explain new avenues to push beyond the status quo and unleash the full power of artificial intelligence's contribution to the central aim of science. We hope our perspective inspires and focuses research towards androids that get new scientific understanding and ultimately bring us closer to true artificial scientists.
\end{abstract}

\maketitle

\section{Introduction}
Artificial Intelligence (A.I.) has recently been called a ``new tool in the box for scientists"\cite{zdeborova2017new} and that ``machine learning with artificial networks is revolutionizing science``\cite{fosel2018reinforcement}. Additionally, it has been conjectured ``that machines could have a significantly more creative role in future research." \cite{melnikov2018active}. For instance, it has even been postulated that ``[t]he new goal of theoretical chemistry should be that of providing access to a chemical 'oracle': an A.I. environment which can help humans solve problems, associated with the fundamental chemical questions of the fourth industrial revolution [...], in a way such that the human cannot distinguish between this and communicating with a human expert" \cite{aspuru2018matter}.

\begin{figure*}[ht]
\centering
\includegraphics[width=0.7\textwidth]{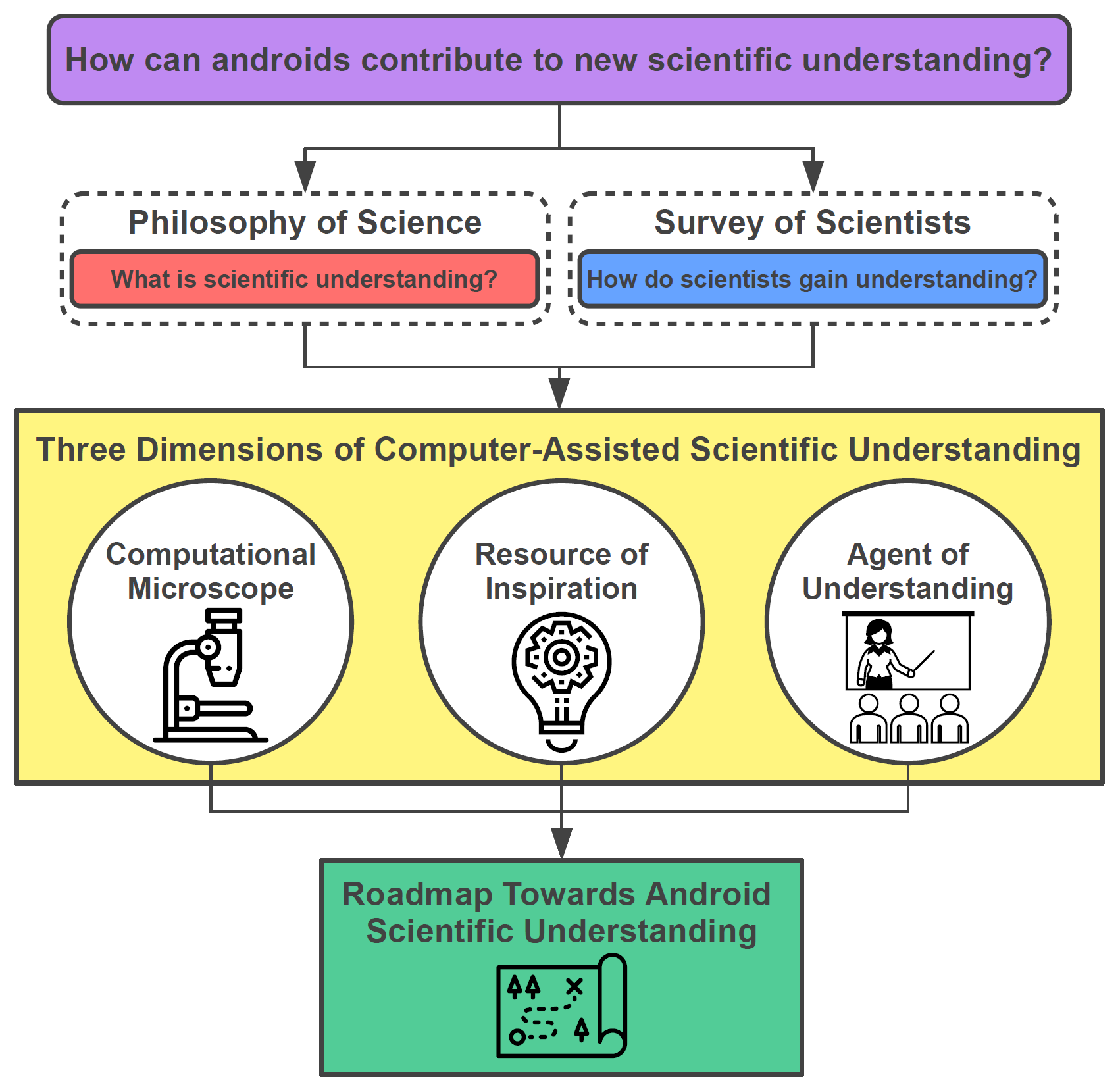}
\caption{\textbf{How can Androids contribute to new scientific understanding?} In addition to scientific literature, we take inspiration from the philosophy of science and from dozens of stories provided by active computational natural scientists. Thereby we identify three fundamental dimensions of computer-assisted scientific understanding. From there, we look into the future and develop a roadmap on how to develop Androids that can contribute to understanding -- the essential aim of science.}
\label{fig:flowofargument1}
\end{figure*}

However, this excitement has not been shared among all scientists. Specifically, it has been questioned whether advanced computational approaches can go beyond \textit{numerics} \cite{hoffmann2020simulationA, hoffmann2020simulationB, hoffmann2020simulationC, marcus2020next,thaler2021} and contribute fundamentally to one of the essential aims of science, that is, gaining of new scientific understanding \cite{potochnik2015diverse,potochnik2017idealization,de2017understanding}.

In this work, we address how artificial systems can contribute to scientific understanding -- specifically, what is the state-of-the-art and how we can push further. Besides a thorough literature review, we surveyed dozens of scientists at the interface of biology, chemistry or physics on the one hand, and artificial intelligence and advanced computational methods. These personal narratives focus on the concrete discovery process of ideas and are a vital augmentation to the scientific literature. We put the literature and personal accounts in the context of a philosophical theory of \textit{Scientific Understanding} recently developed by Dennis Dieks and Henk de Regt \cite{de2005contextual,de2017understanding}, who was awarded the Lakatos Award in 2019 for the development of this theory. We thereby introduce three fundamental dimensions for scientific androids\footnote{We encapsulate all advanced artificial computational systems under \textit{androids}, independent of their working principles. In this way, we are focusing on the operational objective rather than the methodology.} contribution towards new scientific understanding: 
\begin{enumerate}[I)]
\item Androids acting as a microscope in the responses, i.e., akin to an instrument revealing properties of a physical system that are otherwise difficult or even impossible to probe. Humans then lift these insights to scientific understanding.
\item Androids acting as muses, i.e., sources of inspiration for new concepts and ideas that are subsequently understood and generalized by human scientists.
\item Lastly, in an ultimate dimension of android-assisted scientific understanding, computers are the agents of understanding. While we have not found any evidence of computers acting as true agents of understanding in science yet, we outline important characteristics of such an artificial system of the future and potential ways to achieve it.
\end{enumerate}
In the first two dimensions, the android enables humans to gain new scientific understanding while in the last one the machine gains understanding itself. These classes enable us to layout a vibrant and mostly unexplored field of research, which will hopefully manifest itself as a guiding star for future developments of artificial intelligence in the natural sciences.

The goal of this perspective is to put \textit{Scientific Understanding} back to the limelight -- where we are convinced it belongs. We hope to inspire physicists, chemists and biologists and A.I. researchers to go beyond the status quo, focus on these central aims of science, and revolutionize computer-assisted scientific understanding. In that way, we believe that androids will become true agents of understanding that contribute to science in a fundamental and creative way.

\section{Scientific Understanding}
Let us imagine an oracle providing non-trivial predictions that are always true. While such a hypothetical system would have a very significant scientific impact, scientists would not be satisfied. We want ``\textit{to be able to grasp how the predictions are generated, and to develop a feeling for the consequences in concrete situations}" \cite{de2005contextual}. Colloquially, we refer to this goal as ``understanding" -- But what does that really mean? Can we find criteria for scientific understanding? To do that, we seek guidance from the field of philosophy of science. Notably, while hardly any scientist would argue against ``understanding" as an essential aim of science (next to explanation, description and prediction \cite{de2020understanding}), this view was not always accepted by philosophers. Specifically, Carl Hempel, who made foundational contributions clarifying the meaning of \textit{scientific explanation}, argued that ``understanding" is subjective and merely a psychological by-product of scientific activity and is therefore not relevant for the philosophy of science \cite{hempel1965aspects}. Other philosophers criticized these rather unsatisfying conclusions, and they tried to formalize what \textit{scientific understanding} means. Proposals include that understanding is connected to the ability to build causal models (Lord Kelvin said ``It seems to me that the test of 'Do we or not understand a particular subject in physics?' is, 'Can we make a mechanical model of it?' "\cite{de2005contextual}), connected to providing visualizations (or \textit{Anschaulichkeit}, as its strong proponent Erwin Schr\"odinger called it\cite{schrodinger1996nature,de2014visualization}) or that understanding corresponds to providing unification \cite{friedman1974explanation, kitcher1981explanatory}.

In recent years, Henk de Regt and Dennis Dieks have developed a new theory of scientific understanding, which is both contextual and pragmatic \cite{de2005contextual,de2017understanding,de2020understanding}. Importantly, they find that techniques such as visualization or unification are ``tools for understanding", thereby unifying previous ideas in one general framework. Their theory is agnostic to the specific ``tool" being used, making it particularly useful for application in scientific disciplines. They extend crucial insights by Werner Heisenberg \cite{heisenberg1927} and rather than introducing mere theoretical or hypothetical ideas, the main motivation behind their theory is that a ``\textit{satisfactory conception of scientific understanding should reflect the actual (contemporary and historical) practice of Science}". Put simply, they argue that:
\begingroup
\addtolength\leftmargini{-0.1in}
\begin{quote}
A phenomenon P can be understood if there exists an intelligible theory T of P such that scientists can recognise qualitatively characteristic consequences of T without performing exact calculations \cite{de2005contextual,de2017understanding}.
%\end{displayquote}
\end{quote}
\endgroup

Concretely, de Regt and Dieks define two interlinked criteria:
\begin{enumerate}
\item \textbf{Criterion of Understanding Phenomena}: A phenomenon P can be understood if a theory T of P exists that is intelligible.
\item \textbf{Criterion for the Intelligibility of Theories}: A scientific theory T is intelligible for scientists (in context C) if they can recognise qualitatively characteristic consequences of T without performing exact calculations.
\end{enumerate}
%}

We decided to use this specific theory because of one particular strength: We can use it experimentally to evaluate whether scientists have \textit{understood} new concepts or ideas, rather than by inspecting their methodology, by simply looking at the scientific outcome and the consequences. This also coincides with Angelika Potochnik's argument that ``\textit{understanding requires successful mastery, in some sense, of the target of understanding}"\cite{potochnik2017idealization}. We will follow this approach and, consequently, here explore its relationship to the role of A.I. in science. Accordingly, we believe we can significantly advance A.I.'s contribution to this central aim of Science if we have a clear picture of how scientists gain conceptual understanding, and instil it to artificial systems afterwards. We approach this goal by applying ideas of de Regt and Dieks directly to android assisted science (and ultimately, to android scientists themselves).

\section{What is next?}

\subsection{Beyond \textit{Re}-Discovery}
In recent years, scientists at the interface between A.I. and the natural sciences tried to rediscover scientific laws or concepts with machines. The question is, however, whether an android is capable of contributing to new scientific understanding if it can \textit{re}discovers physical laws and concepts, such as the heliocentric world view \cite{iten2020discovering}, the arrow of time \cite{seif2020machine} or mechanical equations of motions \cite{udrescu2020ai}? We believe that this is not guaranteed. The human creators of these androids know what they are looking for in these case studies. Therefore, it is unclear how both conscious and unconscious biases (in the broadest sense, e.g., by choosing particular representations) in the code or the data analysis can be prevented. Consequently, even if an algorithm can rediscover interesting physical phenomena, we cannot know whether and how they can be used to advance Science by helping to uncover new scientific understanding.

Hence, we believe we need to go beyond rediscovery tasks. Therefore we focus explicitly on the question of how to get \textit{new} scientific understanding.

\subsection{Beyond Discovery}
Importantly, other central aims of science such as prediction and discovery can lead to scientific and technological disruptions while not directly contributing to scientific understanding as discussed above \cite{de2020understanding,potochnik2017idealization}. For instance, imagine the hypothetical discovery of the \textit{hitherto} best material for energy storage that could revolutionize batteries. However, this game-changing discovery would not qualify as understanding if chemists could not use the underlying principles fruitfully in other contexts (without computation).

Similarly, the recent breakthrough in protein folding will undoubtedly change the landscape of biochemistry. However, so far, AlphaFold is a black box -- an oracle\cite{jumper2021highly,tunyasuvunakool2021highly}. As such it does not directly provide new scientific understanding in the sense of de Regt and Dieks (but could of course in the future enable humans to gain new scientific understanding). Hence, we believe we must go beyond artificial discoveries in science.

\subsection{Where to go from here?}
The ultimate goal is to get new \textit{understanding} from androids. Loosely speaking, we want to find new ideas or concepts that we can apply and use in different situations without (complete) computations.

This article aims to explain precisely what such a goal requires, what previous approaches have achieved, and how we can go further. We want to clearly lay out this underappreciated but essential research question and thereby give a clear goal for the future of A.I. in the natural sciences.

\begin{figure*}[ht]
\centering
\includegraphics[width=0.85\textwidth]{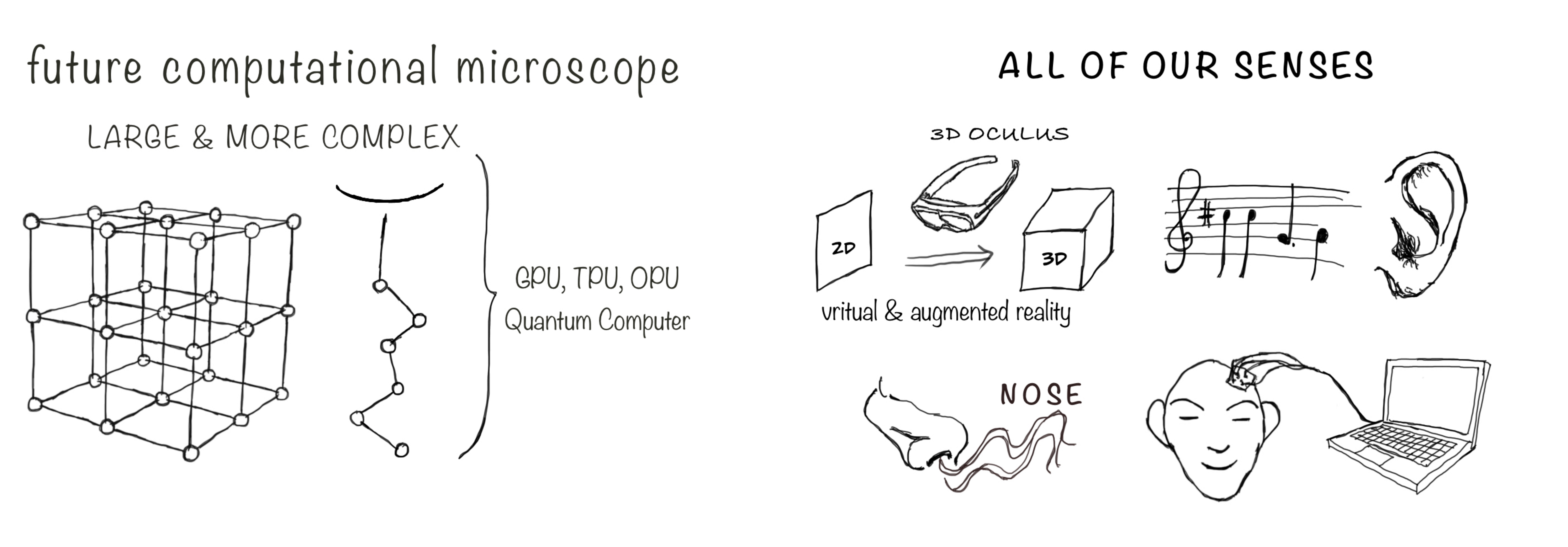}
\caption{\textbf{The future computational microscope.} We envision two types of advances in the next-generation computational microscopes which aim to advance genuine scientific understanding. First (left), larger and more complex computations will allow the computational observation of phenomena not accessible so far. There, new computational paradigms will play a significant role, such as Graphical Processor Units (GPU), Tensor Processor Units (TPU), Optical Processor Units (OPU) and -- ultimately -- quantum computers. Second (right), new ways to represent the highly complex data will advance our ability to sense structure and recognize underlying patterns. The involvement of all our senses could, for sensing computer-generated data, be an exciting pathway to advanced understanding.}
\label{fig:flowofargument2}
\end{figure*}

\section{Three dimensions of computer-assisted understanding}\label{sec:III}
We use scientific literature and personal anecdotes of dozens of scientists, and the context of the philosophy of science, to introduce a new classification of androids contribution to scientific understanding\footnote{We call the classification \textit{dimensions}, as they are independent and non-exclusive.}. It helps to see diverse unexplored journeys that can be investigated in the future.

An android can act
\begin{enumerate}[I)]
\item as a \textbf{computational microscope}, providing information not (yet) attainable by experiment

\item as a \textbf{resource of inspiration} or an \textit{artificial muse}, expanding the scope of human imagination and creativity.
\end{enumerate}

In those two classes, the human scientist is essential to take the new insight and inspiration and develop it to full understanding. Finally, an android can be

\begin{enumerate}[I)]
\setcounter{enumi}{2}
\item an \textbf{agent of understanding}, replacing the human in generalizing observations and transferring scientific concepts to new phenomena. 
\end{enumerate}

We stress that these three classes should not be understood dogmatically but rather guide future possibilities. In the following sections, based on concrete examples, we discuss each class in more detail and propose avenues for pushing the boundaries of the current computational faculties.

\subsection{Computational microscope for scientific understanding}
Microscopes are devices that enable us to investigate objects and phenomena imperceptible to the naked eye. In a similar way, \textit{computational microscopes} enable the investigation of objects or processes that we cannot visualize or probe in any other way. One main objective is to simulate biological, chemical or physical processes that happen at length and time scales not perceivable by experiment.

As we are interested in understanding, the new computer-generated data needs to be generalized to other contexts without complete computation\cite{de2005contextual}. We show now two concrete examples.

The first example is molecular dynamics simulations of the SARS-CoV-2. The authors uncovered new biological functions that show different behaviours in the open and closed conformations of the spike protein. This explanation changed the view upon glycans in biological systems and inspired new ways to analyze these systems without the need to perform full computations \cite{casalino2020beyond}.

In the second example, the authors describe how molecular dynamics simulations helped to uncover fundamental patterns called \textit{glycoblocks}. The systematic use of glycoblocks can both be used to understand sequence-structure-property relationships of biomolecules and can also inform the design of synthetic structures with desired functions without the need for simulating the entire system \cite{fogarty2020and}.

\subsubsection*{The next computational microscope}\label{NextCMicroscope}
A computational microscope aims to provide data via computation that are not (yet) accessible by experiments that humans can understand. How could we make computational microscopes even more insightful and make it easier for human scientists to use this data to gain scientific understanding? There are two vibrant directions going forwards. First, more advanced computational systems will allow to analyze of more complex physical systems. Second, representing the information in a more interpretable way will help to lift the indications from computers to true scientific understanding.

\textit{More Complex Systems} -- One obvious but nevertheless important research direction is increasing the complexity as well as the accuracy of computer simulations\cite{friederich2021machine}. For example, increasing the size of the systems, the time-scale of the simulations, the number of interactions that can be modelled will significantly increase the applicability in complex dynamic systems. In general, this can be achieved by either algorithmic improvements or hardware improvements, or both. In that regard, we expect that modern neural network technologies together with advanced hardware such as GPUs, TPUs or even OPUs \cite{gigan2020artificial,xu202111} will have an enormous impact. Furthermore, the recent progress in experimental quantum computing for quantum chemistry \cite{google2020hartree} and physics\cite{zhang2017observation, schweizer2019floquet, martinez2016real} promises that entirely new algorithms, based on quantum mechanics itself, will play an important role in this area \cite{cao2019quantum,gross2017quantum}. Algorithmic improvements could involve adaptive and intelligent resolution during simulation and advanced visualization methods\cite{de2005contextual}, which directly leads to the second future techniques:

\textit{Full spectrum of senses} -- We believe that human scientists can get more out of data if the full capabilities of all our senses are addressed. At the moment, we analyze data largely in (potentially animated) 2-dimensional pictures. As a first step, we believe that real 3D environments (realized either via virtual or augmented reality glasses, or holography) will significantly help in understanding complex systems or complex data. Initial advances in that regard have been demonstrated in the domain of chemistry \cite{o2018sampling, probst2018exploring, schmid2020structural}, and we expect this to become a standard tool for scientists to advance scientific understanding. In addition, we expect that going beyond the visual sense can open entirely new ways to experience scientific data. For example, the auditory sense is excellent in detecting structure or symmetries in (periodic) time-dependent data \cite{castelvecchi2021sound}. Furthermore, including the sense of touch, smell and taste could further expand the horizon of experiences. We expect that in order to realize that, physical scientists need to work closely together with psychologists and neurologists (and potentially even with artists), to develop suitable data representations that can be efficiently recognized by scientists with all their senses. An ultimate, admittedly futuristic version of a computational microscope could circumvent the receptors of human senses and instead use a computer-brain interface to enhance further experiencing computed data.

\subsection{Resource of inspiration for scientific understanding}
Surprising and creative ideas are the foundation of Science. Computer algorithms are a means to provoke such ideas systematically, thereby significantly accelerating scientific and technological progress. Already 70 years ago,  Alan Turing realized that computers could surprise their human creators: \textit{``Machines take me by surprise with great frequency. This is largely because I do not do sufficient calculation to decide what to expect them to do, or rather because, although I do a calculation, I do it in a hurried, slipshod fashion, taking risks."} and \textit{"Naturally I am often wrong, and the result is a surprise for me for by the time the experiment is done these assumptions have been forgotten``}\cite{turing1950computing}.

A much more recent study provides stories by dozens of researchers of artificial life and evolution. They demonstrate in an impressive way how computer algorithms can surprise their human creators and lead to behaviour that the authors \textit{would denote as creative}\cite{lehman2020surprising}. Accordingly, we believe that androids can be artificial muses of Science in a metaphorical sense.

Those examples demonstrate that computers can indeed be used as a source of surprises. But what are the most general ways to get inspirations from computers? And how can they be lifted by humans to true scientific understanding? We will outline a number of ways to develop ways to provoke surprising behaviours of algorithms and use their solutions, internal or external states as a source of inspiration for new scientific ideas.

\begin{figure*}[ht]
\centering
\includegraphics[width=0.85\textwidth]{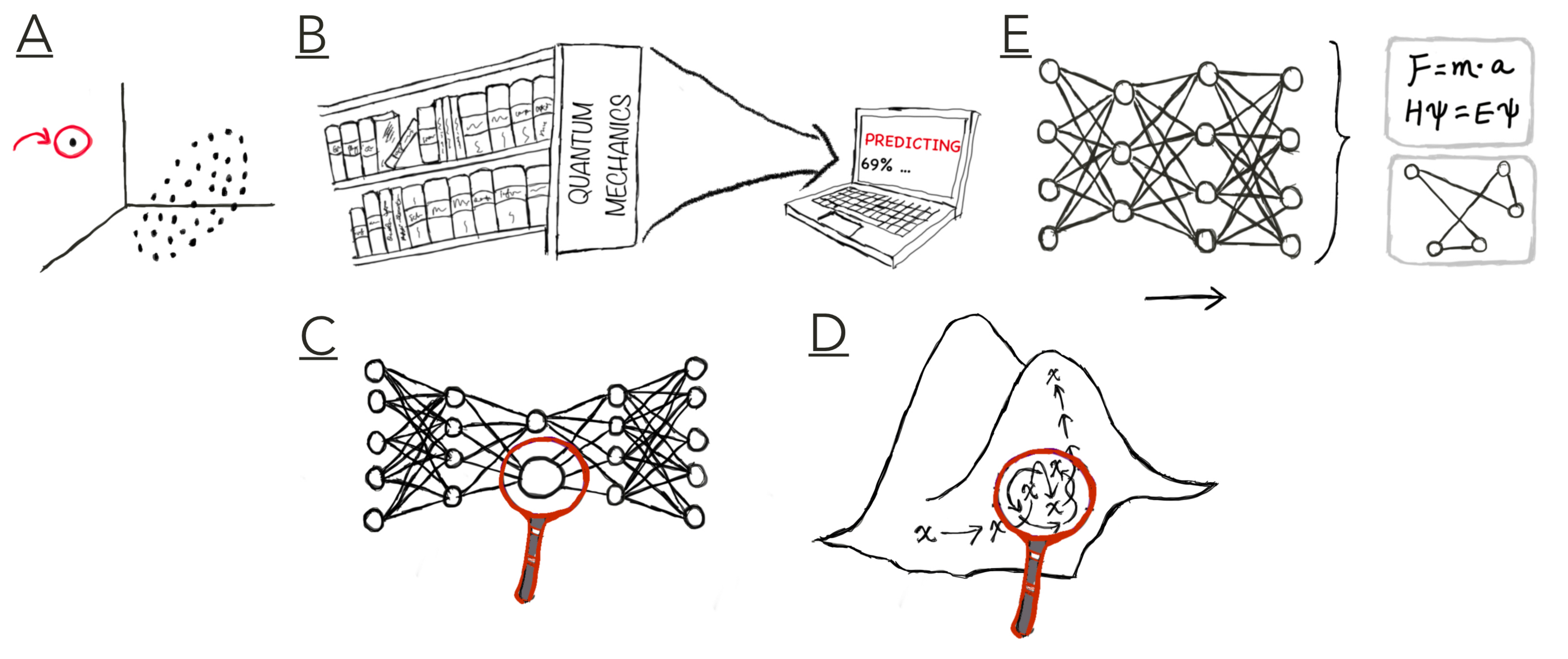}
\caption{\textbf{The future \textit{re}-source of inspiration: } An android could act as a computational muse and inspire the human scientist by (A) identifying surprises in data, (B) identifying surprises in the scientific literature, (C) finding surprising concepts by inspecting models or (D) by probing the behaviour of artificial agents or (E) my finding new concepts from interpretable solutions.}
\label{fig:flowofargument3}
\end{figure*}

\subsubsection*{The future resource of inspiration}
\textit{Identifying surprises in data -- } Exceptional data points or unexpected regularities obtained from experiments or simulations can surprise human scientists and inspire new ideas and concepts. Our survey shows that these exceptional points are usually identified by humans, such as the following two examples, which use high-throughput computations in chemistry \cite{pickard2011ab} and quantum optics \cite{krenn2016automated,krenn2020computer}.

The first example deals with a surprising phase of crystal structures in high-pressure physics. There, the authors found an unexpected stable configurations of alternating \ce{NH2} and \ce{NH4} layers, rather than a dense \ce{NH3} phase. The authors conceptualized this phenomenon as spontaneous ionization, a common process in acid-base chemistry, which is now a widely accepted phenomenon in the high-pressure phase diagram of \ce{NH3}. Spontaneous ionization in the high-pressure behaviour of matter has become a more general principle that can be used without performing any simulations \cite{pickard2008highly}.

In the second example, a search for new quantum experiments uncovered a solution with considerable larger quantum entanglement than expected. The authors understood the underlying principles and thereby discovered a new concept of entanglement generation \cite{krenn2017entanglement, krenn2017quantum}. The principle can be used without any computation and, for example, acts now as a new representation in more advanced artificial intelligence systems for quantum physics \cite{krenn2020conceptual}, demonstrating the application of the computer-inspired idea in more general different contexts.

In contrast to these examples and many others from literature and from personal accounts, the anomalies could manifest themselves in a more involved combination of variables, which might be very difficult for humans to grasp. Accordingly, applying advanced statistical methods and machine learning algorithms (e.g., see reference \cite{malhotra2015long}) to this type of problem will be an important future research direction. Exciting works into the direction of autonomous anomaly detection have been applied on scientific data from the Large Hadron Collider (LHC) at CERN \cite{aad2020dijet, tonon2021probing, park2021quasi}. Such techniques have the potential to identify new physics signatures, which can then be conceptualized and understood by human physicists \cite{schwartz2021modern,kasieczka2021lhc}. Neural networks that autonomously discover symmetries could become an efficient discovery tool for outliers in scientific data where the underlying rules might not be known beforehand \cite{yu2018group,dehmamy2021automatic}.

Estimating the confidence of predictions will be one method to directly search for anomalies in data \cite{nigam2021assigning}. The ability to uncover hidden regularities was very recently demonstrated in mathematics, where an A.I. hinted on relations between previously unconnected invariants in knot theory, which allowed mathematicians to conjecture and prove new theorems \cite{davies2021advancing}. Alternatively, an A.I. capable of constructing new scientific hypotheses could uncover outliers or unexpected patterns that are not discernible with standard statistical methods. 

It would be truly exciting to see an A.I. uncover hidden patterns or irregularities in scientific data previously overlooked by humans, which leads to new ideas and, ultimately, to new conceptual understanding. As of now, we are not aware of cases like that.

The data points for these systems could be obtained from computational methods (involving those described in section \ref{NextCMicroscope}), with exciting opportunities for mathematics or theoretical physics \cite{douglas2022machine}. Alternatively, the data could be obtained directly from experiments. Here we can imagine a closed-loop approach where an algorithm tries to explore the environment and steer the exploration into unexpected regions. If the data-source is an experiment, this future system will require access to complex lab automation with large parameter spaces to explore, as demonstrated recently in biology \cite{king2009automation}, chemistry \cite{bedard2018reconfigurable, steiner2019organic, coley2019robotic, burger2020mobile, chatterjee2020automated, grizou2020curious} or physics \cite{moon2020machine,dalgaard2020global}.

\textit{Identifying surprises in the scientific literature -- } The number of scientific papers in essentially every scientific domain is growing enormously\cite{larsen2010rate}. Consequently, researchers have to specialise in narrow subdisciplines, which makes finding new interdisciplinary ideas difficult. In the future, we believe that computers will be able to use the scientific literature in an automated way\cite{evans2011metaknowledge,clauset2017data,fortunato2018science,wang2021science} and identify exceptional and surprising phenomena for further investigation. While the large-scale automated analysis of the scientific literature, to our knowledge, has not yet been able to induce new scientific understanding, there is significant progress in the field. One promising approach towards this goal is unsupervised word embedding of a large corpus of scientific papers. In that technique, the content of the scientific literature is transformed into a high-dimensional vector space. Recently, this technique has been applied in the domain of material science \cite{tshitoyan2019unsupervised} and rediscovered central scientific concepts such as the periodic table of the elements. Additionally, the results also suggested the existence of previously undiscovered structure-property relationships. Examples include new candidates for thermoelectric materials. Moreover, several other advanced computational techniques are being developed in material science to extract knowledge from the scientific literature and investigate it systematically by A.I. technologies\cite{olivetti2020data}, and can lead to complex scientific conclusions as demonstrated for instance on zeolite transformations \cite{schwalbe2019graph}.

An alternative approach aims to build semantic knowledge networks from large bodies of scientific literature. In these networks, scientific concepts are nodes, and edges carry relational information. In the simplest case, that means two scientific concepts are mentioned in the same scientific paper \cite{rzhetsky2015choosing,krenn2020predicting}. Thus, scientific knowledge is represented as an evolving network, which can be used to identify both islands and unexplored regions of the scientific literature. This type of network was used in biochemistry to identify efficient global research strategies \cite{rzhetsky2015choosing} and in quantum physics to predict and suggest future research directions \cite{krenn2020predicting}. Advances in A.I. technology could improve this type of system significantly. For example, natural language processing architectures such as BERT \cite{devlin2018bert}, or GPT3 \cite{brown2020language} could help extract more scientific knowledge from research papers, and large graph-based neural networks could improve the prediction of new research topics from semantic networks \cite{hamilton2017inductive}. 

\textit{Surprising concepts by inspecting models --} We also expect considerable progress by rationalising what A.I. algorithms have learned in order to solve a specific problem, i.e., explainable or interpretable A.I. \cite{montavon2018methods,samek2019explainable,roscher2020explainable,lundberg2020local}. One idea towards this goal is inspired by DeepDreaming, a method first used in computer vision \cite{mahendran2015understanding,mordvintsev2015inceptionism}. Put simply, the idea is to invert a neural network and probe its behaviour. Recently, this approach has been applied to rediscover thermodynamical properties \cite{seif2020machine}, and design principles for functional molecules \cite{shen2020deep}. An alternative and remarkable application is the \textit{disentanglement of variables} in neural networks \cite{burgess2018understanding}. The goal is to understand the internal representation the neural network has learned. Recently, astronomical data, represented in geocentric coordinates, was used to train a neural network and disentanglement of variables enabled the rediscovery of heliocentric coordinates via the internal representation of the model \cite{iten2020discovering}. In a related study, using gradient boosting with decision trees, feature importance has been used to explain properties of molecules, and quantum optics circuits \cite{friederich2021scientific}. Related to this is a study where the internal representation of an unsupervised deep generative model for quantum experiments has been inspected to understand the model's internal worldview\cite{flam2021learning}. In the chemical domain, counterfactual explanations for machine learning models have been demonstrated to produce rationale behind a model's prediction. Counterfactual explanations illustrate what differences to an event or instance would generate a change in an outcome. Wellawatte et al. \cite{wellawatte2022model} showed how this can be achieved in a model-independent way (it has been demonstrated for random forest, sequence models and graph neural networks), indicating great future potentials for opening the black-box of AI in science. Albeit not in science, recent work has investigated what the chess-playing A.I AlphaZero has learned about chess and how human-like knowledge is encoded in the internal representation \cite{mcgrath2021acquisition}. The concepts rediscovered in all of those works were not new, and thus the most important challenge for the future is to learn how to extract previously unknown concepts. Progress towards resolving that challenge will be essential in the near future to inspire new scientific ideas.

\textit{New concepts from interpretable solutions --} Rather than getting inspirations from the A.I. algorithms themselves, scientists can also be surprised by the corresponding solutions. When solutions are represented in an interpretable way, they can provoke new ideas and lead to new concepts. An example of interpretable representations is a mathematical formula. Thus, scientists can inspect formulae derived by computer algorithms to solve mathematical problems directly and derive more general solution strategies. Several publications demonstrated extracting symbolic models from experimental data of mechanical systems \cite{schmidt2009distilling, udrescu2020ai}, of quantum systems \cite{gentile2021learning} and in astronomy \cite{cranmer2020discovering}. It will be exciting to see how these approaches, e.g., combined with methods such as causal inference \cite{cranmer2020frontier}, can be improved to propose reasonable physical models of unknown systems that advance scientific understanding. Altogether, exciting advances have been achieved in the field of mathematics\cite{raayoni2021generating,wagner2021constructions}, and we foresee similar approaches making a significant impact in the physical sciences as well. 

One concrete, recent example in astronomy is the rediscovery of Newton's law of gravitation from real-world observational data of planets and moons in our solar system from the last 30 years \cite{lemos2022rediscovering}. The application of graph neural networks allowed for the high-quality prediction of the object's motion. Furthermore, a symbolic regression technique called PySR (introduced in \cite{cranmer2020discovering}) was able to extract reasonable mathematical expressions for the learned behaviour. Interestingly, besides the equations of motions, the method simultaneously predicts the masses of the planetary objects correctly. The technique required the assumption of several symmetries and other physical laws. It will be interesting to see whether these prerequisites can be reduced further and how related approaches can be applied to modern physics questions.

Another concrete example of this methodology has been showcased in the field of quantum optics \cite{krenn2020conceptual}. There, an A.I. algorithm with a graph-theoretical representation of quantum optical setups designs configurations for previously unknown quantum systems. The final solutions were represented in a physically-interpretable graph-theoretical representation. From there, human scientists can quickly interpret the underlying reasons why the solutions work and apply it in other contexts without further computation. Accordingly, developing interpretable representations and methods to extract underlying concepts in other domains will be an important future research direction.

\textit{Probing the behaviour of artificial agents --} Another only rarely explored opportunity is interpreting the behaviour of machines when tasked to solve a scientific problem \cite{rahwan2019machine}. Algorithms that take actions such as genetic algorithms or reinforcement learning agents adopt policies to navigate the problem space. Human scientists can observe how they navigate this space. Instead of following a strict external reward, e.g., maximise a specific property of a physical system, intrinsic rewards such as artificial curiosity can be implemented \cite{schmidhuber2008driven,pathak2017curiosity}. Instead of maximizing directly some functions, the artificial agent tries to learn and predict the behaviour of the environment. It then chooses actions that lead to situations it cannot predict well, thus maximizing its own understanding of the environment. It has been shown using curious agents in simulated virtual universes \cite{thiede2020curiosity} and robot agents in real laboratories \cite{grizou2020curious} that curiosity is an efficient exploration strategy. Alternative intrinsic rewards for artificial agents are \textit{computational creativity}\cite{varshney2020explaining,varshney2019big} and \textit{surprise} \cite{itti2009bayesian}. These intrinsic rewards can produce exceptional and unexpected solutions, ultimately inspiring human scientists.

\subsection{Agent of Understanding}
The third and final class we consider are algorithms that can autonomously acquire new scientific understanding, a feat that has neither been described by the respondents of our survey nor in the scientific literature. Therefore, we will approach this class by listing the requirements of these agents, proposing tests to detect their successful realization and speculating what such computer programs could look like.

First, it is important to realize that finding \textit{new} scientific understanding is context-dependent. What is new depends on whether we consider an individual scientist and their field of expertise, a scientific domain, the whole scientific community or even the entire scientific endeavour throughout history. Hence, true agents of understanding must be able to evaluate whether an insight is new, at least in the context of a specific scientific domain that requires access to the knowledge of a scientific field.

Secondly, de Regt emphasized the importance of underlying scientific theories that allow us to recognize qualitatively characteristic consequences \cite{de2017understanding}. It is not enough to simply interpolate data points or predict new ones using advanced statistical methods such as machine learning. Thus, even though such methods can approximate complex and expensive computations, na\"ive applications of neural networks cannot be agents of understanding. Scientific understanding requires more than mere calculation. To illustrate this point even further, let us consider one concrete example in quantum physics from the literature: A computational method solved an open question about the generation of important resource states for quantum computing. Then it extracted the conceptual core of the solution in the form of a new quantum interference effect in such a fashion that human scientists can both understand the results and apply the acquired understanding in different contexts \cite{krenn2020conceptual}. Even if the computer itself was able to apply the conceptual core to other situations, it would not be \textit{a priori} clear whether the computer truly acquired scientific understanding. What is still missing is an explanation of the discovered technique in the context of a scientific theory. In this particular example, the android and the human scientist would need to recognize the underlying quantum interference in the context of the theory of quantum physics. Thus, we can propose the first sufficient condition for agents of understanding:

\begingroup
\addtolength\leftmargini{-0.17in}
\begin{quote}
\textbf{Condition for Scientific Understanding I:}
\textit{An android gained scientific understanding if it can recognize qualitatively characteristic consequences of a theory without performing exact computations and use them in a new context}.
\end{quote}
\endgroup

This condition closely follows the ideas of de Regt and Dieks \cite{de2005contextual}. Let us go one step further and imagine that there is an android capable of explaining discoveries in the context of scientific theories. How could human scientists recognize that the machine acquired new scientific understanding? We argue that human scientists would do it in the exact same way they can recognize that other human scientists acquired new scientific understanding. That is, let the other human scientists transfer the newly acquired understanding to themselves. This allows us to propose the second sufficient condition for agents of understanding:

\begingroup
\addtolength\leftmargini{-0.17in}
\begin{quote}
\textbf{Condition for Scientific Understanding II:}
\textit{An android gained scientific understanding if it can transfer its understanding to a human expert}.
\end{quote}
\endgroup

We argue that one can only recognize indirectly whether a computer (or human) has gained scientific understanding. Therefore, finally, we propose a test in the spirit of the Turing test \cite{turing1950computing} or the Feigenbaum test\cite{feigenbaum2003some} (or adaptations thereof in the natural sciences such as the Chemical Turing Test or the Feynman Test \cite{aspuru2018matter}):

\begingroup
\addtolength\leftmargini{-0.1in}
\begin{quote}
\textbf{The Scientific Understanding Test:}

\textit{A human (the student) interacts with a teacher, either a human or an android scientist. The teacher's goal is to explain a scientific theory and its qualitative, characteristic consequences to the student. Another human (the referee) tests both the student and the teacher independently\footnote{In principle, there is no reason for the student or the referee not to be androids. However, to keep the test as simple as possible, we want to keep the number of possible variations small.}. If the referee cannot distinguish between the qualities of their non-trivial explanations in various contexts, we argue that the teacher has scientific understanding.}
\end{quote}
\endgroup

This implies that \textit{humans} need to understand the new concepts that androids devised. If a machine truly understands something, it will be able to explain it and transfer the understanding to someone else.\footnote{We leave aside the question whether the explanation of the android is true or false. It has been argued that also false theories can lead to genuine understanding \cite{de2017false}.} We believe that this should always be possible, even if the understanding is far beyond what human experts know at this point. We envision that computers will use advanced human-computer interaction techniques together with the tools we described for (future) computational microscopes.

Additionally, scientific discussions between a human and a computer could be realized using advanced queries in natural language processing tools such as BIRD \cite{devlin2018bert} or GPT-3 \cite{brown2020language}. That way, the scientist could probe the computer with scientific questions. Suppose the scientist gains new scientific understanding by communicating with the algorithm, as judged by our scientific understanding test. In that case, they can confirm that the computer truly acquired understanding.\footnote{We would like to point out that our test, like the ones originated by Turing and Feigenbaum, are not clear-cut, leaving room for situations that do not allow a clear judgement.} We are optimistic that more efforts will be directed at developing the necessary technologies, which will lead to ever more convincing demonstrations of android scientists acting as true agents of understanding in the future.

\section{Conclusion}
Undoubtedly, advanced computational methods in general and artificial intelligence specifically will further revolutionize how scientists investigate the secrets of our world. We outline how these new methods can directly contribute to one of the main aims of science, namely acquiring new scientific understanding. We suspect that significant future progress in the use of androids to acquire scientific understanding will require multidisciplinary collaborations between natural scientists, computer scientists and philosophers of science. Thus, we firmly believe that these research efforts can -- within our lifetimes -- transform androids into true agents of understanding that will directly contribute to one of the most essential aims of science, namely Scientific Understanding.

\section*{Acknowledgements}
The authors thank Anastassia Alexandrova, Rommie Amaro, Curtis Berlinguette, Lillian Chong, Gerardo Cisneros, Andy Cooper, Graeme Day, Francois-Xavier Coudert, Lee Cronin, Elisa Fadda, Rafael Gomez-Bombarelli, Leticia Gonzalez, Johannes Hachmann, Roald Hoffmann, Jan Halborg Jensen, Erin R. Johnson, Lynn Kamerlin, Heather J. Kulik, Jean-Paul Malrieu, Anat Milo, Frank Noe, Jens Kehlet N{\o}rskov, Artem Oganov, Juan Perez-Mercader, Chris Pickard, Markus Reiher, Jean-Louis Reymond, Dennis Salahub, Stefano Sanvito, Franziska Schoenebeck, Ilja Siepmann, Alex Sodt, Isaac Tamblyn, Donald Truhlar, Alexandre Tkatchenko, Koji Tsuda, Alexandre Varnek, Tejs Vegge, Anatole von Lilienfeld and Eva Zurek for answering our questions on understanding, Xuemei Gu for Figure 2 and 3, and Nora Tischler and Robert Fickler for valueable comments on the manuscript. A.A.-G. and his group acknowledge generous support from the Canada 150 Research Chairs Program, the University of Toronto, and Anders G. Fr\o seth. M.K. acknowledges support from the FWF (Austrian Science Fund) via the Erwin Schr\"odinger fellowship No. J4309. R.P. acknowledges funding through a Postdoc.Mobility fellowship by the Swiss National Science Foundation (SNSF, Project No. 191127).

\bibliography{refs}

\end{document}